\documentclass{aa} 
\usepackage{graphicx}
\usepackage{epsfig}
\begin{document}
   \title{The complex X--ray spectrum of NGC~4507}

   \author{Giorgio Matt
          \inst{1}
          \and
          Stefano Bianchi\inst{1}
	  \and
	  Filippo D'Ammando\inst{1}
	  \and
 	  Andrea Martocchia\inst{2}
          }

   \offprints{G. Matt}

   \institute{Dipartimento di Fisica, Universit\`a degli Studi Roma Tre,
via della Vasca Navale 84, I-00146 Roma, Italy
         \and
Observatoire Astronomique, 11 Rue de l'Universit\'e, 
F--67000 Strasbourg, France
\\
             }

   \date{Received ; accepted }

   \abstract{XMM--$Newton$ and $Chandra$/HETG spectra of the Compton--thin 
($N_{\rm H}\sim4\times10^{23}$ cm$^{-2}$) Seyfert 2 galaxy, NGC~4507, 
are analyzed and discussed. 
The main results are: a) the soft X--ray emission is rich in emission lines;
an (at least) two--zone photoionization region is required to explain the large range of
ionization states. b) The 6.4 keV iron line is likely emitted from Compton--thick matter,
implying the presence of two circumnuclear cold regions, one Compton--thick (the emitter),
one Compton--thin (the cold absorber). c) Evidence of an Fe {\sc xxv} absorption line is found
in the $Chandra$/HETG spectrum. The column density of the ionized absorber is estimated to be
a few$\times10^{22}$ cm$^{-2}$.
               }

   \maketitle
%

\section{Introduction}

NGC~4507 is a nearby ($z$=0.0118) spiral galaxy and one of the X--ray brightest 
Compton--thin Seyfert 2s, despite the heavy obscuration 
(N$_{\rm H}\sim4\times10^{23}$ cm$^{-2}$). 
{\sl ASCA} clearly detected, in addition to the absorbed power and the iron K$\alpha$ line
already observed by {\sl GINGA} (Awaki et al. 1991), a strong soft X--ray excess and an
intense emission line at $\sim$0.9 keV (Comastri et al. 1997; Turner et al. 1997), 
identified as the Ne {\sc ix} recombination line. $Beppo$SAX observed the source three times
(Risaliti 2002); in the third observation the flux was about half that of the previous
two observations. A Compton reflection component was also clearly 
detected; interestingly, its relative normalization
about doubled in the third observation. This suggests a constant reflection flux in spite of
the primary flux variations, and therefore an origin in distant matter. 

In this paper we present results from XMM--$Newton$ and $Chandra$/HETG observations of NGC~4507.
In Secs.~2 and 3 data reduction and analysis are described, while the results are summarized
and discussed in Sec.~4.

In the following, we will assume $H_0$=70 km/s/Mpc. 

\section{Data reduction}

\subsection{XMM--Newton}

NGC~4507 was observed by XMM-\textit{Newton} between January 4 and 5, 2001, with the EPIC CCD 
cameras, the pn (Str\"uder et al. 2001) and the MOS (Turner et al. 2001), both operating in 
Full Frame, but adopting Medium and Thick filters, respectively. We will only use pn data in 
this paper, whose observed count rate is much lower than the maximum defined for a 1\% pileup 
(see Table 3 of the XMM-\textit{Newton} Users' Handbook). Following the procedure suggested in 
the \textsc{SAS} 
website\footnote{http://xmm.vilspa.esa.es/sas/documentation/threads/\\PN$\_$spectrum$\_$thread.html}, 
we screened for intervals of flaring particle background by inspecting the high energy (E $>$ 
10 keV) light curve. The resulting net exposure time is 32 ks. X-ray events corresponding to 
pattern 0 were used for the pn together with an extraction radius of $40\arcsec$. The data were 
reduced with \textsc{SAS} 5.4.1.

In this paper, we do not discuss RGS data, because the source is too faint in the RGS energy
range to provide useful spectra.

\subsection{Chandra}

\textit{Chandra} observed the source on March 15 2001 for 140 ks, with the ACIS-S HETG. 
Unfortunately, the default 3.2 s frame time resulted in a high degree of pileup for the zeroth 
order spectrum. Therefore, we will only use the HEG and MEG 1st order spectra, rebinned  
to have at least 50 counts per bin. Data were reduced with \textsc{ciao} 3.0.1 and 
\textsc{caldb} 2.23, following standard procedures.\\

\section{Data analysis}

All spectra were analysed with \textsc{Xspec} 11.2.0. In the following, errors correspond
to the 90\% confidence level for one interesting parameter ($\Delta \chi^2 =2.71$).

\begin{table}[t]
\begin{tabular}{|l|c|}
\hline
& \cr
$\Gamma_{\rm soft}$ & 3.07$^{+0.15}_{-0.22}$  \cr
& \cr
$\Gamma_{\rm hard}$ & 1.80$^{+0.14}_{-0.17}$  \cr
& \cr
N$_H$ (10$^{23}$ cm$^{-2}$) & 4.40$^{+0.54}_{-0.57}$ \cr
& \cr
F(1 keV)$_{\rm soft}$/F(1 keV)$_{\rm hard}$ & 0.015 \cr
& \cr
R & 1.5 (fixed) \cr
& \cr
$\chi^2$/d.o.f. & 218.7/175 \cr
& \cr
\hline
\end{tabular}
\caption{XMM--$Newton$ spectrum: best fit results for the continuum. See text for details}
\label{fit}
\end{table}

\subsection{XMM--Newton}

\begin{figure}
\epsfig{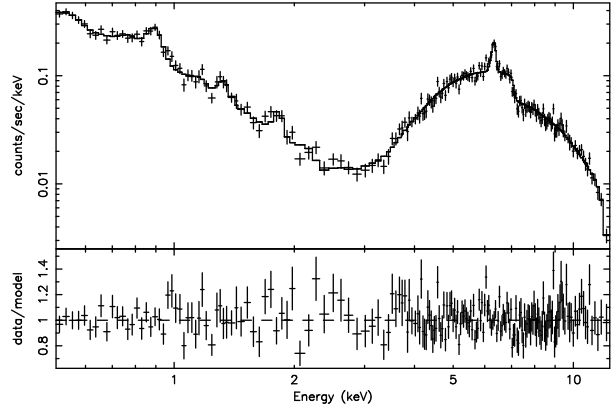}
\caption{XMM--$Netwon$ spectrum and best fit model, and data/model ratio. See text for details.}
\label{bestfit}
\end{figure}

Following Comastri et al. (1997), 
we fitted the XMM--$Newton$ spectrum of NGC~4507 with a model composed of
an absorbed power law plus Compton reflection continuum ({\sc pexrav} model in XSPEC; the 
inclination angle fixed to 30$^{\circ}$, the metal abundances to the solar value
as in Anders \& Grevesse 1989);
a soft, unabsorbed (apart from Galactic absorption, N$_{H,gal}=7.19\times10^{20}$ cm$^{-2}$) 
power law;  several emission lines to fit the features clearly present
in the spectrum (see Fig.~\ref{bestfit}). All absorption and emission features (apart from
the Galactic absorption) have been corrected for the source redshift.
The results are summarized in Tables~\ref{fit} and \ref{fit_lines}.
Because of the limited bandwidth of XMM--$Newton$, which prevents an accurate determination
of the Compton reflection component, we fixed $R$ (i.e. the relative normalization
between reflected and primary components) to 1.5, which is the value expected
(on the basis of $Beppo$SAX results, Risaliti 2002)
at the flux level of the source during the XMM--$Newton$ observation (i.e. 1.28$\times10^{-11}$
erg~cm$^{-2}$~s$^{-1}$ in the 2--10 keV band, corresponding to an unabsorbed luminosity
of 1.5$\times10^{43}$ erg~s$^{-1}$) under the hypothesis that the 
reflection arises from distant matter and it is therefore constant despite source flux
variations. (Leaving $R$ free to vary, it is very loosely constrained).

Several emission lines are clearly detected. Besides confirming the presence of the iron
K$\alpha$ and the 0.9 keV lines, already discovered by ASCA (Comastri et al. 1997), many
other lines are found (Table~\ref{fit_lines}).  Apart from the 6.4 keV iron line, all
other unambiguously detected lines come from He--like ions, with the exception of a
faint (and marginally detected) line from H--like neon. 
Possible lines from sulphur are also present, but they are in an energy region, i.e. around
2.5 keV, that is notoriously affected by calibration problems for the pn.
The rather broad ionization structure
can hardly be explained by just a single, homogenous zone (Netzer \& Turner 1997, Guainazzi et 
al. 1999): at least two photoionized regions (one for the oxygen and neon lines, the other for
the magnesium and silicon lines), are required.

Evidence for the Fe K$\alpha$ Compton shoulder (CS; see Matt 2002 and references therein)
is also found. The Compton shoulder is significant at the 99.6\% confidence level,
according to the F-test. 
The large ratio between the CS and the line core, which is greater than
0.2, suggests Compton-thick matter. This, together with the significant Compton
reflection component observed by $Beppo$SAX (Risaliti 2002), suggests that the bulk of the
line does not originate in the Compton--thin absorber, which therefore should have a small
covering factor (see e.g. Matt et al. 2003). This finding provides one more argument in favour
of the presence of both Compton--thin and Compton--thick circumnuclear regions in AGN
(Maiolino \& Rieke 1995; Matt 2000; Weaver 2002; Matt et al. 2003).

No Fe K$\beta$ line is detected. The ratio between the upper limit to the Fe K$\beta$ flux
and the  Fe K$\alpha$ total (i.e. core + Compton Shoulder) flux is $<$12\%, suggesting
that iron is more ionized than {\sc xii} (but less than {\sc xvii} from the line energy;
see discussion in Molendi et al. 2003).

No absorption lines are required by the data. At low energies, this is not
surprising, because  the nuclear radiation is not directly visible due to the heavy
cold absorption, and the observed emission is therefore due to reflection.   
No evidence for a 6.7 keV  absorption line is found (see next section). The upper
limit to the flux (keeping the energy and width fixed to the values found in the 
$Chandra$/HETG spectrum, see next section), 
is --1.3$\times10^{-5}$ ($\sigma$=0) or --1.6$\times10^{-5}$
($\sigma$=0.1 keV) ph~cm$^{-2}$~s$^{-1}$. Because the flux in the XMM--$Newton$ 
observation was about a factor of 2 lower than in the $Chandra$ observation, these upper limits
are consistent with constant absorption properties.

\begin{table*}[t]
\begin{tabular}{|c|c|c|c|c|}
\hline
& & & &\cr
E (keV) & F (10$^{-6}$ ph cm$^{-2}$ s$^{-1}$) & EW (eV) & Identification & F({\it Chandra})  \cr
& & & &\cr
\hline
& & & & \cr
0.571$^{+0.011}_{-0.016}$ & 50.1$^{+12.1}_{-17.0}$ & 47 & O {\sc vii} K$\alpha$ & $<$340 \cr
& & & &\cr
0.65 (fixed) & 0$^{+5.3}_{-0}$ & 0 & O {\sc viii} K${\alpha}$ & $<$140 \cr
& & & &\cr
0.784$^{+0.050}_{-0.024}$ & 9.0$^{+3.4}_{-5.6}$ & 22 & O {\sc vii} RRC, O {\sc viii} K$\beta$,
Fe {\sc xvii} L & $<$38 \cr
& & & &\cr
0.912$^{+0.008}_{-0.010}$ & 24.1$^{+3.9}_{-4.1}$ & 92 & Ne {\sc ix} K${\alpha}$  & $<$26 \cr
& & & &\cr
1.176$^{+0.038}_{-0.048}$ & 1.9$^{+1.8}_{-1.7}$ & 16 & Ne {\sc x} K${\alpha}$ & $<$4.5 \cr
& & & &\cr
1.344$^{+0.018}_{-0.017}$ & 4.1$^{+1.6}_{-1.6}$ & 52 & Mg {\sc xi} K${\alpha}$ & $<$5 \cr
& & & &\cr
1.843$^{+0.038}_{-0.063}$ & 2.6$^{+1.1}_{-1.2}$ & 85 & Si {\sc xiii} K${\alpha}$ & $<$2.3 \cr
& & & &\cr
6.32 (fixed)  & 23.2$^{+17.5}_{-11.4}$ & 47 & Fe {\sc i--xvi} CS & 22$^{+37}_{-19}$  \cr
& & & &\cr
6.403$^{+0.021}_{-0.028}$ & 59.0$^{+14.1}_{-13.6}$ & 117 & Fe {\sc i--xvi} K${\alpha}$ &
99$^{+29}_{-25}$ \cr
& & & &\cr
7.06 (fixed) & 0$^{+6.9}_{-0}$ & 0 & Fe {\sc i--xvi} K${\beta}$ & $<$13 \cr
& & & &\cr
\hline
\end{tabular}
\caption{XMM--$Newton$ spectrum: best fit results for the emission lines. 
All EWs are calculated against the soft
power law, apart for the iron K lines which are calculated against the  hard power
law.  See text for further details.  In the last column, the line fluxes from
the $Chandra$ observation are given.}  
\label{fit_lines}
\end{table*}

\begin{figure}
\epsfig{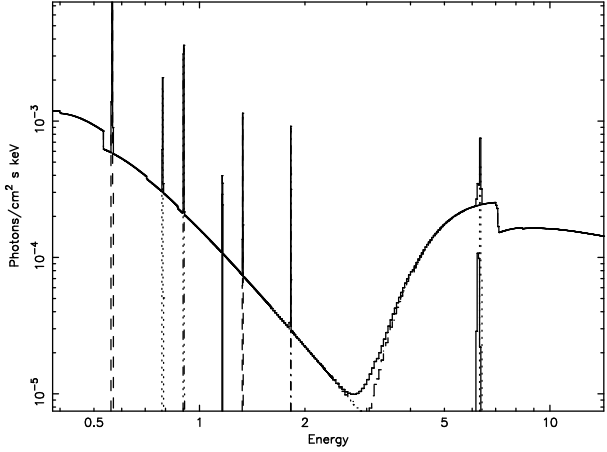}
\caption{The unfolded best fit spectral model for the XMM--$Netwon$ spectrum.}
\label{model}
\end{figure}

\subsection{Chandra}

During the $Chandra$ observation the 2--10 keV source 
flux was 2.37$\times10^{-11}$ erg cm$^{-2}$ s$^{-1}$,
i.e. about a factor 2 higher than in the XMM--$Newton$ observation. Because of the modest 
signal--to--noise in the continuum, we limited the analysis to the 4--8 keV band, to exploit
the HETG energy resolution at the iron line energies (no other emission line is 
detected, upper limits to the fluxes being consistent with XMM--$Newton$ measurements;
see Table~\ref{fit_lines}).

\begin{figure}
\epsfig{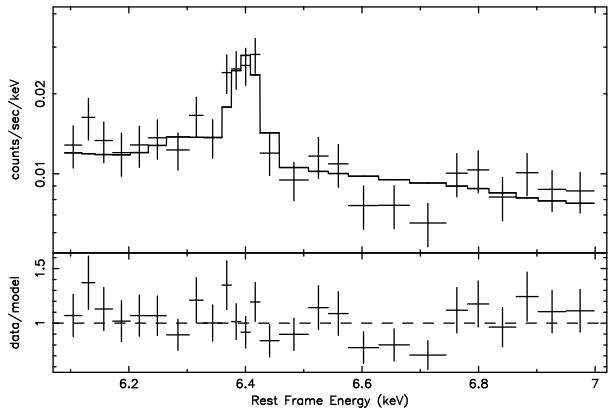}
\caption{$Chandra$/HETG spectrum and best fit model, and data/model ratio, around the
iron lines energies.}
\label{absline}
\end{figure}

The spectrum was fitted with an absorbed power law ($\Gamma$ and $N_{\rm H}$ fixed to the best
fit values for the XMM--$Newton$ spectrum), plus the iron K$\alpha$ line (core plus CS, 
see previous section). The fit is good ($\chi^2$/d.o.f.=60.3/87), but an absorption
feature around 6.7 keV is apparent in the residuals (see Fig.~\ref{absline}). We fitted it
with a Gaussian line with negative normalization; the $\chi^2$/d.o.f. becomes 52.8/84, and  
the line is significant at the 98.9\% confidence level, according to the F-test.

The fluxes of the core and CS of the iron emission line 
are 9.9$^{+2.9}_{-2.5}\times10^{-5}$ ph cm$^{-2}$ s$^{-1}$
and 2.2$^{+3.7}_{-1.9}\times10^{-5}$ ph cm$^{-2}$ s$^{-1}$, respectively. The line flux
is therefore marginally consistent with being constant between the XMM--$Newton$ and $Chandra$
observations (see Table~\ref{fit_lines}) but, given the large errors especially for the 
$Chandra$ results, the possibility that the iron line flux follows continuum variations
cannot be excluded, either. The core of the line is unresolved, with an
upper limit of 26 eV to $\sigma$.

The centroid energy of the absorption line is 
6.68$^{+0.03}_{-0.04}$ keV, the upper limit to $\sigma$ is 0.1 keV. Both line energy and
width are therefore consistent with the blend of the Fe {\sc xxv} recombination triplet.
The line flux is --(3.0$\pm$1.8)$\times10^{-5}$ ph cm$^{-2}$ s$^{-1}$ (EW=--26 eV). 

Finally, no evidence for extended emission is found, either in the soft or
in the hard bands. This means that the emitting regions are
smaller than about 800 kpc (corresponding to 1'' at the source distance).

\section{Discussion}

\subsection{Soft X--ray continuum and line emission.}

The soft X--ray emission is well fitted  by a steep ($\Gamma\sim$3) power law plus several
emission lines from oxygen, neon, magnesium and silicon. The ionization state implied by
the observed lines is too wide for a single photoionized region, or for
collisionally ionized matter. At least two photoionized reflecting
regions (or perhaps a region with a 
range of ionization parameters, as in NGC~1068, see Kinkhabwala et al. 2002) is
therefore required. If the soft X--ray emission is indeed due to reflection of the nuclear
radiation by ionized matter, this implies a significant steepening of the intrinsic emission
at low energies,
not unusual in Seyfert galaxies (see e.g. Vaughan et al., 2004, and references therein).

The He--like to H--like K$\alpha$ lines ratio for both oxygen and neon is very large, i.e.
at least a factor of 10 (see Table~\ref{fit_lines}). The He--like line is actually a triplet,
consisting of resonance, intercombination and forbidden lines, while the H--like line 
is a doublet of resonance lines. The resonance lines can easily be optically thick, with
consequent reduction of their equivalent width due to resonant trapping (Matt et al. 
1996). This is not true for the intercombination and, especially, forbidden lines which have
much lower oscillator strengths: for low densities the
forbidden line may indeed dominate the triplet, as shown by Porquet \& Dubau (2000).

The line energies are consistent with rest frame values: no strong outflow is 
required by the data. The geometry of the reflecting regions is hard to derive with
the present data. These regions must lie outside the cold,
Compton--thin absorber in order to be observable. The lack of any extended emission
seen by $Chandra$ implies a not very constraining upper limit to the size of the
reflecting regions of about 800 pc.

\subsection{The cold reprocessor}

The iron 6.4  keV emission line is clearly detected, along with its Compton shoulder. 
The shoulder--to--core ratio clearly points toward Compton--thick matter (Matt 2002), as did
the earlier $Beppo$SAX detection of a Compton reflection continuum (Risaliti 2002).
The constancy of the reflection component flux despite significant primary flux
variations and the narrowness of the iron line as measured by $Chandra$ suggest
that the Compton--thick matter is rather distant from the black hole, and likely to be
identified with the molecular torus. This in turn implies the presence of two
different cold circumnuclear regions, one Compton--thick (the reflector) the other Compton--thin
(the absorber), as it is now often found in Seyfert 2s (Matt et al. 2003, and references
therein). The Compton--thin region should have a rather low covering factor (as
seen by the nucleus) so as not to overproduce the iron line.

\subsection{The ionized Fe {\sc xxv} absorption line}

In the $Chandra$/HETG spectrum, evidence for a  Fe {\sc xxv} absorption line was found.
The line is significant at about a 99\% confidence level. The ionization state of the absorbing
matter is much higher than that of the photoionized regions responsible for the soft X--ray lines.

Ionized iron absorption lines have been already observed in several sources. In some cases
the line energies are significantly blueshifted, implying large velocity and very massive
outflows (e.g. Pounds et al. 2003; Reeves et al. 2003, and references therein). In other
cases (Reeves et al. 2004; Vaughan \& Fabian 2004), iron absorption occurs in matter with
no detectable motion, as in our case. It is very likely that the two absorbing regions
are completely different, the first probably being much closer to the black hole and possibly
due to a disk wind (e.g. King \& Pounds 2003; Proga et al. 2000), the second likely 
related to the `hot' scattering regions often observed (via reflected continuum and ionized
iron emission lines) in Seyfert 2 galaxies (the best studied
case is NGC~1068, Matt et al. 2004 and references therein). Recently, such lines have also 
been found in
unobscured Seyferts, thanks to the improved sensitivity of XMM--$Newton$ (Matt et al. 2001;
Bianchi et al. 2003a).  The matter responsible for these lines
is possibly located much further away than the outflowing matter, and indeed 
in one case, NGC~5506 (where highly ionized iron emission lines were observed), 
it is likely extended over several hundreds of parsecs (Bianchi
et al. 2003b); however, in the case of NGC~3783 the variability of the line depth
argues in favour of a much smaller (less than 0.1 pc) size (Reeves et al. 2004). 
The line EW in NGC~4507 corresponds (assuming solar iron abundance, an ionization state
peaked on He--like iron and moderate turbulence to ensure the matter is optically
thin in the line; see Nicastro, Fiore \& Matt, 1999, for the relevant formulae and graphs) 
to an absorbing hydrogen equivalent column density of a few $\times10^{22}$ cm$^{-2}$, i.e. 
of the same order as the one estimated by Bianchi \& Matt (2002) for NGC~5506.

\section*{Acknowledgements}
Based on observations obtained with XMM--$Newton$, an ESA science mission with 
instruments and contributions directly funded by ESA Member States and the USA
(NASA). GM and SB acknowledge financial support from ASI and MIUR (under grant  
{\sc cofin--03--02--23}).

\end{document}